\documentclass[ERTS, 2026]{ERTS} 

\usepackage{graphicx}
\usepackage{amsmath}

\makeatletter
\AddToHook{cmd/maketitle/before}{\let\thetitle\@title}
\makeatother

\usepackage{amssymb,xspace}
\usepackage{mathtools}
\usepackage{algorithm,hyperref}
\usepackage{subfig}
\usepackage{algorithm}
\usepackage[beginLComment=/*~, endLComment=~*/]{algpseudocodex}
\algrenewcommand\algorithmicfunction{\textbf{Function:}}
\algrenewcommand\algorithmicoutput{\textbf{Output:}}

\newcommand{\acetone}{{\sc Acetone}\xspace}

\usepackage[disable]{todonotes}
\usepackage{todonotes}
\setuptodonotes{fancyline, inline, color=blue!30}

\begin{document}

\title{Extension of \acetone C code generator for multi-core architectures}

\author{%
	Yanis Aït-Aïssa\authorNumber{1,2}, Thomas Carle\authorNumber{3}, Sergeï Chichin\authorNumber{2}, Benjamin Lesage\authorNumber{1} and Claire Pagetti\authorNumber{1}\\ 
    $^1$ ONERA, $^2$ Airbus, $^3$ Université de Toulouse –  IRIT -- all in Toulouse France
}

\address{}

\maketitle

\chead{\thetitle}

\pagestyle{fancy}

\thispagestyle{plain}

\licenseFootnote{Yanis Aït-Aïssa~et al}

\begin{abstract}
    Integrating Deep Neural Networks into safety-critical aeronautical systems remains a significant challenge, as it requires high predictability and adherence to precise certification standards. The ACETONE framework addresses this by generating certifiable C code for mono-core systems. We present an extension to the framework for the generation of predictable, parallel C code on multi-core architectures. By modeling the inference as a DAG scheduling problem and utilizing an optimized Constraint Programming approach, we enable efficient execution on multi-core CPUs without dedicated accelerators. Our solution is validated through both WCET analysis and experimental evaluation.
\end{abstract}

\todo{Pour désactiver les TODO avant la soumission, décommentez dans le fichier \texttt{main.tex} l'appel au package \emph{todonotes} qui contient l'option disable. }

\section{Introduction}
The use of Machine Learning in industry has grown steadily in recent years. In the aeronautical sector, it is an appealing way forward for enabling new functionalities such as navigation assistance, runway detection, and trajectory optimization. However, integrating such advanced algorithms in safety-critical systems remains a significant challenge. 
This has led to the emergence of new development assurance processes, as outlined in the 
EASA guidance \cite{Easaconcept}
and the yet to be published draft of the ED 324, with publicly available material \cite{gabreau:hal-WG,kaakai2022toward,kaakai2023datacentric}.
 In this work, we focus on the implementation and inference of deep neural networks (DNN) on multi-core architectures. 
 Indeed, the aeronautical sector is currently moving from single core to multi-core and is not yet ready to embed dedicated accelerators \cite{lesage25}. 
Thus, for a rapid solution, it 
makes sense to have a multi-core-based execution
for small- and medium-sized networks.

\subsection{Context}

The framework \acetone \cite{dealbuquerquesilva_et_al:LIPIcs.ECRTS.2022.3} tackles 
part of this problem by generating sequential C code from detailed descriptions of off-line trained DNNs.
The generated code is designed to be easily certifiable for being embedded in aeronautical safety-critical systems. 
Practically, the code preserves the semantics of the DNN model by 
1) coding the operations described in the input's description 
and 2) assuring that 
the C output numerical values on  data
is very close to those computed by the training framework (in the range of the numerical error). 
The code is also predictable, which means that its worst-case execution time (WCET) can be estimated, and this was shown with the OTAWA \cite{otawa} tool. 
However, the framework is limited to run DNN inference on the same core,
which
quickly increases the computation time and in turn may breach the time constraints given by the application. Parallelizing the C code on multi-core architectures 
allows reducing the global inference time, and thus exploiting more time constrained or even larger models. 
However, the size of the board's memory remains a limiting factor with respect to
the model size.

The transition from single-core to multi-core is far from trivial. In a theoretical perspective, parallelizing code on a multi-core platform requires dividing the application into independent computational blocks, each of which is assigned to a core for execution with the goal of minimizing the overall execution time. 
From an implementation perspective, programming for multi-core processors requires the use of \emph{synchronization} mechanisms to coordinate the various cores.
From \acetone perspective, all the internal generation processes are optimized to produce sequential code.


\subsection{Contributions}
In this paper, we address both the theoretical and practical aspects of the 
generation of parallel C code in the field of embedded systems for aeronautics
and propose an extension to \acetone. 
Our contribution is threefold:
\begin{itemize}
    \item We formalize the computation of an offline parallel schedule as a DAG scheduling problem. To this end, we define the embedded system, the application model representing the studied neural networks, and the expected schedule.
    \item To compute (optimal) schedules, we encode the search as a constraint programming problem, similar to Tang \textit{et al.} \cite{TANG2020115}. 
   Such encoding scales poorly, leading us to 
   propose a much more efficient equivalent problem
   (by reducing the number of decision variables and improving some of the constraints). 
   We then
   review efficient heuristics \cite{Kruatrachue}.
    \item We extend \acetone to generate parallel code. 
    As the experiments are deployed in a bare-metal setting,
    we integrate  within \acetone both the mapping of programs to cores and adequate 
    \emph{synchronization} mechanisms.
    
    \item We evaluate our predictable extension both by computing the WCET obtained with  OTAWA \cite{otawa} and by measuring the observed WCET for various neural networks.
\end{itemize}

The outline of the paper is as follows. Section \ref{sec-pbm-statement} details the problem statement.
Section \ref{sec-dag-schedule} describes the offline DAG  scheduling problem and some ways of computing (optimal) solutions.
Section \ref{sec-evaluation} presents the comparison of those solutions.
Section \ref{sec-acetone-extension} focuses on the extension of \acetone.
We finally conclude the paper and present some future work in Section \ref{sec-conclusion}.

\section{Problem statement}
\label{sec-pbm-statement}
\subsection{Platform model}

The platform consists of a multi-core CPU with a finite number $m$ of identical cores operating under a unified memory architecture (UMA). The homogeneity of the cores ensures that the WCET of a task is independent of the core on which it is executed. As a result, the WCET of a sequential code can be computed on any core, with identical outcomes. The UMA allows each core to access the shared memory at the same speed and ensures that inter-core communication latency does not vary with the source and destination. However, such memory representation induces interference between the cores during concurrent memory access. The precise study of interference is considered out of scope for our study and is represented by a margin added to all computed WCET. 



Parallelism is implemented using a shared memory strategy, meaning that each core can read from and write to a common memory space. These operations are not instantaneous and introduce latency. In particular, a source core writes a message to a predetermined memory location, and the destination core reads it once a synchronization mechanism permits.

\subsection{Application model}
\label{application_model}

\begin{figure}[hbt]
    \centering
    \includegraphics[width=\linewidth]{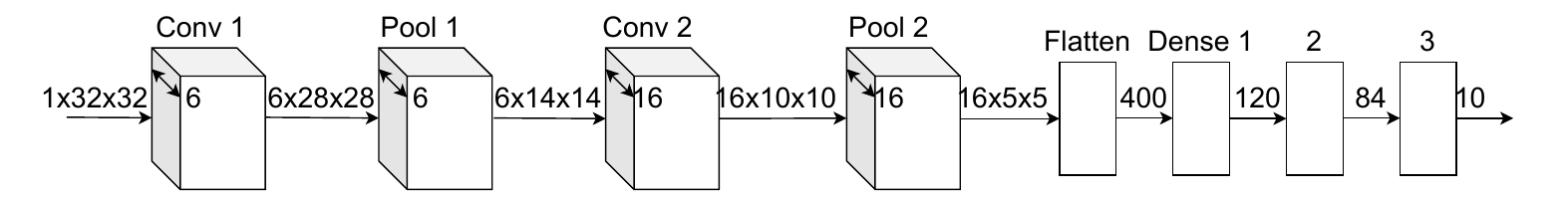}
    \caption{LeNet-5 neural network}
    \label{fig:lenet}
\end{figure}

An application is defined as a deep neural network (DNN). The framework extension relates to the code generated and not the type of input; the DNN is assumed to be compatible with \acetone, meaning the network is an off-line pre-trained feed-forward network. In particular, the network is assumed to be either a convolutional neural network (CNN) -- such as the LeNet-5 neural network presented in Figure \ref{fig:lenet} -- or simply a multilayer perceptron (MLP), also referred to as a fully connected neural network.

In this work, we focus on the scheduling 
and code generation of a given neural network with a fixed architecture. Practically, each layer appearing in the DNN description is translated into a sequential C code.
This allows to keep the code generated by \acetone of each layer untouched and to preserve the good properties required by certification and offered by \acetone.

\begin{figure}[hbt]
    \centering
    \includegraphics[width=\linewidth]{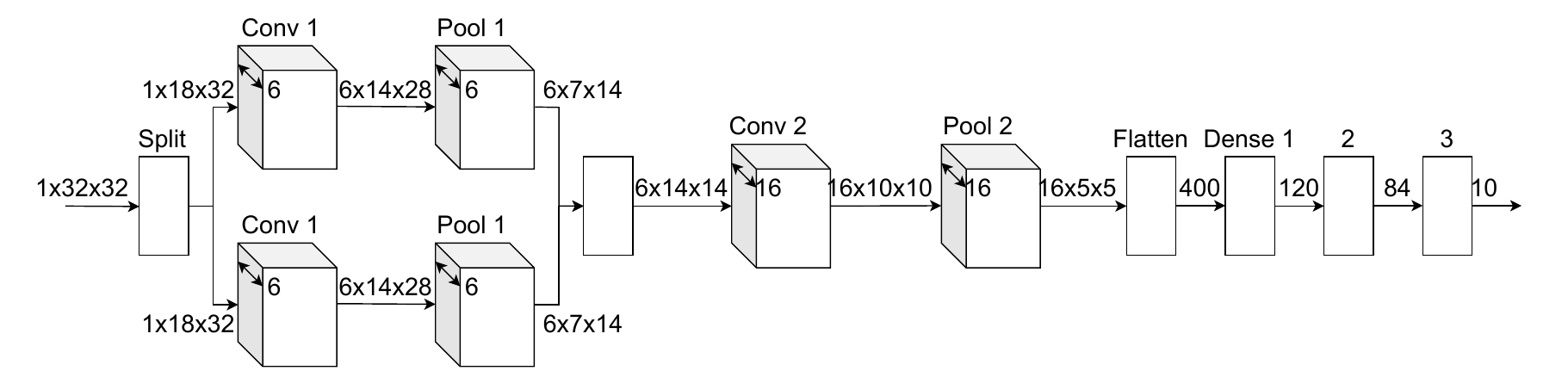}
    \caption{Modified LeNet-5 neural network}
    \label{fig:lenet_sep}
\end{figure}

If needed, the user can modify the initial neural network beforehand to obtain a more parallel architecture. For example, the LeNet-5 in Figure \ref{fig:lenet} has a purely sequential architecture and is not adapted for parallelization. It can be, for instance, transformed by splitting 
the first two layers, as done in \cite{gauffriau2024formaldescriptionmlmodels} and shown in Figure \ref{fig:lenet_sep}, to create two separate parallel branches. 
Such preliminary off-line modifications to expose high parallelism are left to the user, and our work solely focuses on scheduling the input model without modifying its architecture. 



The DNN architecture is modeled as a directed acyclic graph (DAG), where each layer of the network is associated with a node of the graph. A directed graph (DG) is defined by a quadruplet $(V,E,t,w)$, where:
\begin{itemize}
    \item $V$ is the set of the graph's nodes. Each node represents a layer from the network.
    \item $E \subset V \times V$ is the set of the graph's edges. For two distinct nodes $u$ and $v$ from $V$, the edge $e=(u,v) \in E$ if and only if there is an exchange of data from $u$ to $v$. $u$ is then said to be a parent of $v$, and $v$ is a child of $u$.
    \item $t: V  \mapsto \mathbb{R}$ is a function mapping to each node the WCET of the corresponding task on a core.
    \item $w: E \mapsto \mathbb{R}$ is a function mapping to each edge the latency that would be induced by the transmission of data from the source of the edge to its destination, if both of them are not on the same core.
\end{itemize}
A DG is said to be acyclic if and only if there is no succession of nodes linked by edges from a node to itself. 

\begin{figure}[hbt]
    \centering
    \includegraphics[width=0.7\linewidth]{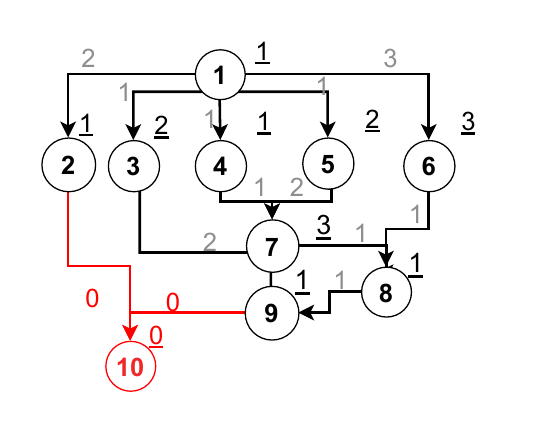}
    \caption{Example of a DAG. In black the original graph, and in red the new node and edges used to compute a one-sink graph.}
    \label{fig:dag}
\end{figure}

We assume that DAGs have only one sink node denoted subsequently $s$. It is easy to transform any DAG into an equivalent one sink DAG. 
The black nodes in Figure \ref{fig:dag} illustrate an DAG with 9 nodes. Each node is identified by a unique label, while the WCET is indicated by the underlined values on its right. The transmission delay on an edge is indicated by the gray value next to it. 
The red part illustrates the transformation into a one sink DAG.

\subsection{Model expression}
We assimilate a node from the graph to the corresponding task in the schedule. 
The scheduling is done in \acetone before generating the code and abides by the following properties:
\begin{enumerate}
    \item The scheduling is done statically.
    \item The tasks are non-preemptive, i.e. 
    once they have started their execution, they do not stop until completion. 
    \item A task can be scheduled if and only if all of its predecessors have been entirely scheduled -- the task is then said to be ready.
\end{enumerate}

A schedule is a tuple $(\mathcal{S}c_1,\ldots,\mathcal{S}c_m)$, where each $\mathcal{S}c_i$ represents a sub-schedule for core $i$. A sub-schedule $\mathcal{S}c_i$ is defined as a list of tuples $(v, s)$, where $v \in V$ is a node of the graph and $s \in \mathbb{R}$ denotes the start time of $v$. This list is empty if no nodes are assigned to the corresponding core.

A schedule is said to be valid if the following properties hold:
\begin{itemize}
    \item Two nodes cannot be executed at the same time on the same core.
    \item A node does not start before its inputs are available (the parents have been executed, and the communication is over).
    \item Each node can be duplicated on different cores. Indeed, such duplication can help minimize the makespan by removing the communication delay induced by data transfer between cores. 
    A duplication providing no gain is called \emph{redundant} and is to be removed.
    \item Each node must be present at least once in the schedule and at most once by sub-schedule. 
\end{itemize}
The goal is to find valid schedules with the shortest makespan.

\section{Optimal schedule search}
\label{sec-dag-schedule}
The problem was studied and proved to be NP-complete by Garey and Johnson \cite{Garey}. As such, numerous studies have worked on exact and heuristic searches of the problem.

\subsection{State of the art ILP encoding}
Tang \textit{et al.} \cite{TANG2020115} proposed a constraints representation and used an ILP solver to find optimal solutions. 
The model is based on four decision variables:  
\begin{itemize}
  \item $s_{v_p}$: the start time of task $v \in V$ on core $p \in \mathbf{P}$.
  \item $f_{v_p}$: the completion time of task $v \in V$ on core $p \in \mathbf{P}$.
  \item $x_{v_p}$: a binary variable equal to 1 if node $v \in V$ is scheduled on core $p \in \mathbf{P}$ and is not redundant, and 0 otherwise.
  \item $d_{a_i,b_j}$: a binary variable equal to 1 if task $a \in V$ scheduled on core $i \in \mathbf{P}$ transfers data to node $b \in V$ scheduled on core $j \in \mathbf{P}$, and 0 otherwise.
\end{itemize}

The first constraint ensures that each node is scheduled at least once.
\begin{equation}
    \sum_{p \in \mathbf{P}} x_{v_p} \geq 1, \forall v \in V
    \label{eq:instance}
\end{equation}

Since the tasks are non-preemptive, the completion time equals the start time plus the WCET of the task.
\begin{equation}
    f_{v_p} = s_{v_p} + t(v) \times x_{v_p}, \forall v \in V, \forall p \in \mathbf{P}
    \label{eq:fin}
\end{equation}

When a task is not assigned to a core, its corresponding start time is set to 0.
\begin{equation}
    x_{v_p} = 0 \Rightarrow s_{v_p} = 0, \forall v \in V, \forall p \in \mathbf{P}
    \label{eq:start}
\end{equation}
Thus, by applying \ref{eq:fin} and \ref{eq:start}, we can deduce that when a node is not assigned to a core, its start time and end time on that core are both equal to zero.

A core can execute at most one task at a given instant.
When two tasks are scheduled on the same core, the start time of one must necessarily be after the completion time of the other.
\begin{equation}
    \begin{split}
        x_{a_i}=x_{b_i}=1 \Rightarrow (f_{a_i} \leq s_{b_i}) \vee (f_{b_i} \leq s_{a_i}),\\
        \forall (a,b)\in V\times V \backslash\{a\}, \forall i\in \mathbf{P}
    \end{split}
    \label{eq:contention}
\end{equation}

For each edge $e = (a,b) \in E$, if $a$ is scheduled on core $i \in \mathbf{P}$  and $b$ on $j \in \mathbf{P}$, 
the start time of $b$ cannot be earlier than the completion time of $a$, plus the communication delay needed to transmit data from $a$ to $b$ (if applicable).
\begin{equation}
    \begin{split}
        d_{a_i,b_j} = 1 \Rightarrow f_{a_i} + (1-\mathbf{1}_{i,j}) \times w(e) \leq s_{b_j},\\
         \forall (a,b) \in E, \forall (i,j) \in \mathbf{P}^2
    \end{split}
    \label{eq:prece}
\end{equation}
where $\mathbf{1}_{i,j}$ is the indicator function equal to 1 if $i = j$, and 0 otherwise.

To limit the number of duplications, each node in the schedule must either produce or consume data. 
Indeed, duplicating tasks is of interest to avoid communicating time. Thus, sink nodes are never duplicated.
\begin{equation}
    \sum_{p \in \mathbf{P}} x_{s_p} = 1 \textit{ with $s$ the sink node of the graph}
    \label{eq:leaves}
\end{equation}

In addition, constraint \ref{eq:sender} ensures that every source node of a communication sends data to at least one destination. In other words, for a node $a \in V$ on core $i \in \mathbf{P}$, there exists at least one non-zero variable $d_{a_i,b_j}$, with $b \in V$ and $j \in \mathbf{P}$.
\begin{equation}
    x_{a_i} = 1 \Rightarrow \sum_{b\in\mathcal{S}(a)}\sum_{j\in\mathbf{P}}d_{a_i,b_j} \geq 1, \forall a \in V, \forall i \in \mathbf{P}
    \label{eq:sender}
\end{equation}

Constraint \ref{eq:receiver} ensures that each scheduled destination node of a communication receives all the required data from exactly one source. In other words, for a node $b \in V$ on core $j \in \mathbf{P}$, only one communication variable $d_{a_i,b_j}$ is nonzero for each of its parent nodes $a \in V$.
\begin{equation}
    x_{b_j} = 1 \Rightarrow \sum_{i\in\mathbf{P}}d_{a_i,b_j} = 1, \forall (a,b) \in E, \forall j \in \mathbf{P}
    \label{eq:receiver}
\end{equation}

\subsection{Improved encoding}
Tang's ILP representation can be quite time-consuming. Indeed, the communication variable $d_{a_i,b_j}$ is represented by a 4D tensor and can be complex for the solver to handle. To reduce this complexity, we introduce a new formulation which reworks all the constraints involving $d_{a_i,b_j}$ (constraints \ref{eq:prece}, \ref{eq:sender}, and \ref{eq:receiver}), and
uses only the other three decision variables. All other constraints remain untouched.

To replace constraints \ref{eq:sender} and \ref{eq:receiver} and thereby limit the number of duplications, an upper bound, imposed by constraint \ref{eq:upper_bound_dupli}, prevents a node from having more instances than child nodes.
\begin{equation}
    \sum_{i \in \mathbf{P}} x_{v_i} \leq card(\mathcal{S}(v)), \forall v \in V \backslash \{s\}
    \label{eq:upper_bound_dupli}
\end{equation}
where $card(\mathcal{S}(v))$ is the number of children of node $v$, and $s$ is the sink node of the graph.  
This upper bound on the number of duplications comes from the fact that, for each non-sink node of the graph, an assignment in the schedule is useful if and only if it sends data to another node. Moreover, each node can receive data from only one instance of each parent. Thus, if a node is duplicated more times than it has children, some instances will not communicate with any node.

Constraint \ref{eq:prece}, which introduces the communication delay in the representation, is transformed into two new constraints: if both nodes of an edge $(a,b) \in E$ are assigned to the same core, then the end time of the source must be less than or equal to the start time of the destination (constraint \ref{eq:prec_same_core}). Otherwise, the start time of the destination must occur after the earliest finish time among all instances of the source, plus the communication latency (constraint \ref{eq:prec_diff_core}).
\begin{equation}
    x_{u_i}=x_{v_i}=1 \Rightarrow f_{u_i} \leq s_{v_i}, \forall (u,v)\in E, \forall (i,j )\in \mathbf{P}²
    \label{eq:prec_same_core}
\end{equation}
\begin{equation}
    \begin{split}
        (x_{u_i}=0) \wedge (x_{v_i}=1) \Rightarrow earliest\_f_u  + w(e)\leq s_{v_i},\\
        \forall e=(u,v)\in E, \forall (i,j )\in \mathbf{P}²       
    \end{split}
    \label{eq:prec_diff_core}
\end{equation}
where $earliest\_f_u = \min_{i \in \mathbf{P}} f_{u_i}$ is the earliest end time of an assigned node $u$.

When determining the earliest end time of a node $u$, a conflict may arise with constraint \ref{eq:fin}: if the node is not assigned to a core $i$, the end time $f_{u_i}$ is set to 0, which is the lowest value it can attain and interferes with the earliest end time of the node. To resolve this issue, the constraint is split into two parts: if the node $v \in V$ is assigned to core $i \in \mathbf{P}$, the end time is equal to the start time plus the node's WCET (constraint \ref{eq:fin_wcet}); otherwise, the end time is set to the theoretical maximum, defined as the sum of all nodes' WCETs (constraint \ref{eq:fin_top}).

\begin{equation}
    x_{v_i} = 1 \Rightarrow f_{v_i} = s_{v_i} + t(v), \forall v \in V, \forall i \in \mathbf{P} 
    \label{eq:fin_wcet}
\end{equation}

\begin{equation}
    x_{v_i} = 0 \Rightarrow f_{v_i} = \sum_{u\in V} t(u), \forall v \in V, \forall i \in \mathbf{P} 
    \label{eq:fin_top}
\end{equation}

Such a representation is efficient up to several hundred nodes, which is more than enough to schedule the largest models we can encounter. However, this number can be reached and exceeded when we use finer parallelization. Indeed, the operation behind some layers, such as the convolution layer, can be divided into smaller operations, increasing the number of tasks to be scheduled. Heuristics may become necessary to schedule a finer representation of neural networks.

\subsection{Scheduling heuristics}
\label{sh}

A first possible method is to use a list scheduling algorithm. This family of algorithms has been widely studied in the literature under various constraints and is known to offer high performance. Among them, the most effective are the \textit{critical path algorithms}, which are based on assigning a level to each node—this level determines the order in which nodes are considered during the scheduling phase.

Kruatachue \cite{Kruatrachue} extended these algorithms to a problem similar to ours: new scheduling heuristics applied to a DAG with non-unit tasks and non-zero transmission times. Those heuristics share the same framework: each node is assigned a level equal to the sum of all node execution times alongside the longest valid path from the node to the leaf; then the ready nodes are placed in an ordered list (sorted by level). As long as not all nodes in the DAG are scheduled, the algorithm updates the ready queue (by adding the newly ready tasks to the queue and ordering it by level), selects the node at the front of the queue, finds the core that minimizes the task's start time, and then assigns it accordingly. 

In the first heuristic, Insertion Scheduling Heuristic (ISH), each node is assigned to the core that minimizes the task's start time. An insertion step then places the node on the selected core. If there is an idle period on the core between the previously scheduled task and the current one, the step checks the ready queue and attempts to schedule a node with a lower level in this idle time, while preserving the current task's start time.

\begin{figure}[hbt]
    \centering
    \includegraphics[width=0.8\linewidth]{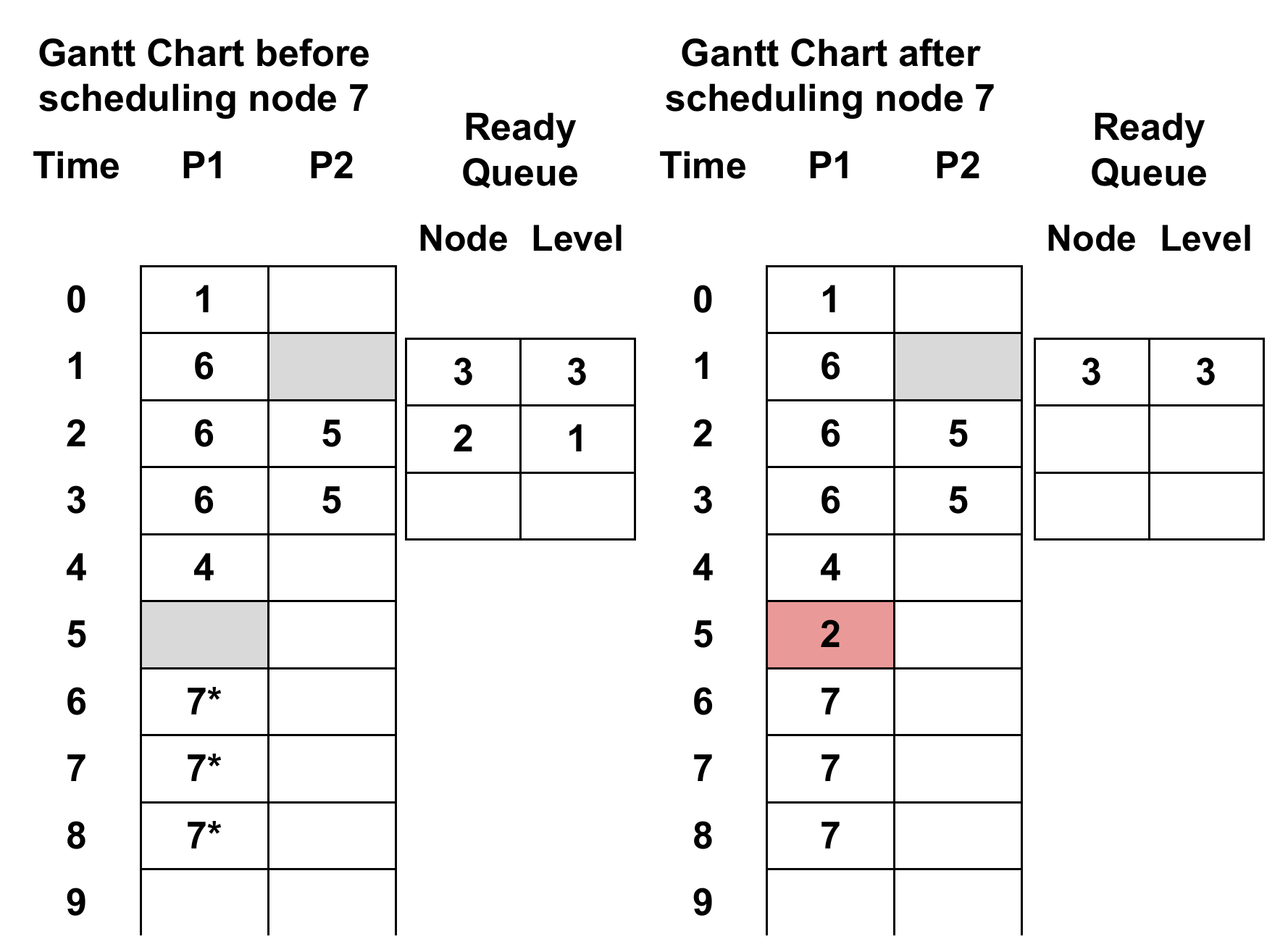}
    \caption{Scheduling of the node 7 using ISH. A "*" means that the node is not yet scheduled, a gray background is a delay induced by the communication between cores and a red background means the node has been scheduled during the insertion step.}
    \label{fig:ish}
\end{figure}

Figure \ref{fig:ish} illustrates this algorithm when trying to schedule the graph from Figure \ref{fig:dag} on two cores. When scheduling node 7, the earliest start time possible is at time 6, due to the communication delay with node 5. This delay creates an idle time of the core — shown in gray in the figure — between times 5 and 6, as shown by the rightmost Gantt chart. This idle time, in turn, triggers an insertion step. Two nodes, 3 and 2, are in the ready queue. Node 3 is parsed first, but the idle time is shorter than its WCET, the second node is considered. Node 2's WCET is short enough, and since it has no communication delay (its only parent, node 1, is on the same core), it fits in this idle time and is inserted into the gap, as shown in the leftmost Gantt chart in Figure \ref{fig:ish}.

In the second heuristic, Duplication Scheduling Heuristic (DSH), most of the complexity lies in the first part. Instead of directly using the start time of a task on the core, the heuristic attempts to optimize that start time. Specifically, if idle time occurs when scheduling a task — most often due to a transmission delay — the function tries to duplicate the task's parent(s) on that core and checks whether the idle time is reduced. If not, it proceeds to duplicate the parents of those parents, and so on, until either no more predecessor(s) remain to duplicate, in which case the process is abandoned, or the  original task's start time is improved, in which case a list of nodes to duplicate on the core is constructed. The second step is similar to that of the previous heuristic, with the additional action of scheduling the duplicated nodes on the appropriate core.

\begin{figure}[hbt]
    \centering
    \includegraphics[width=0.8\linewidth]{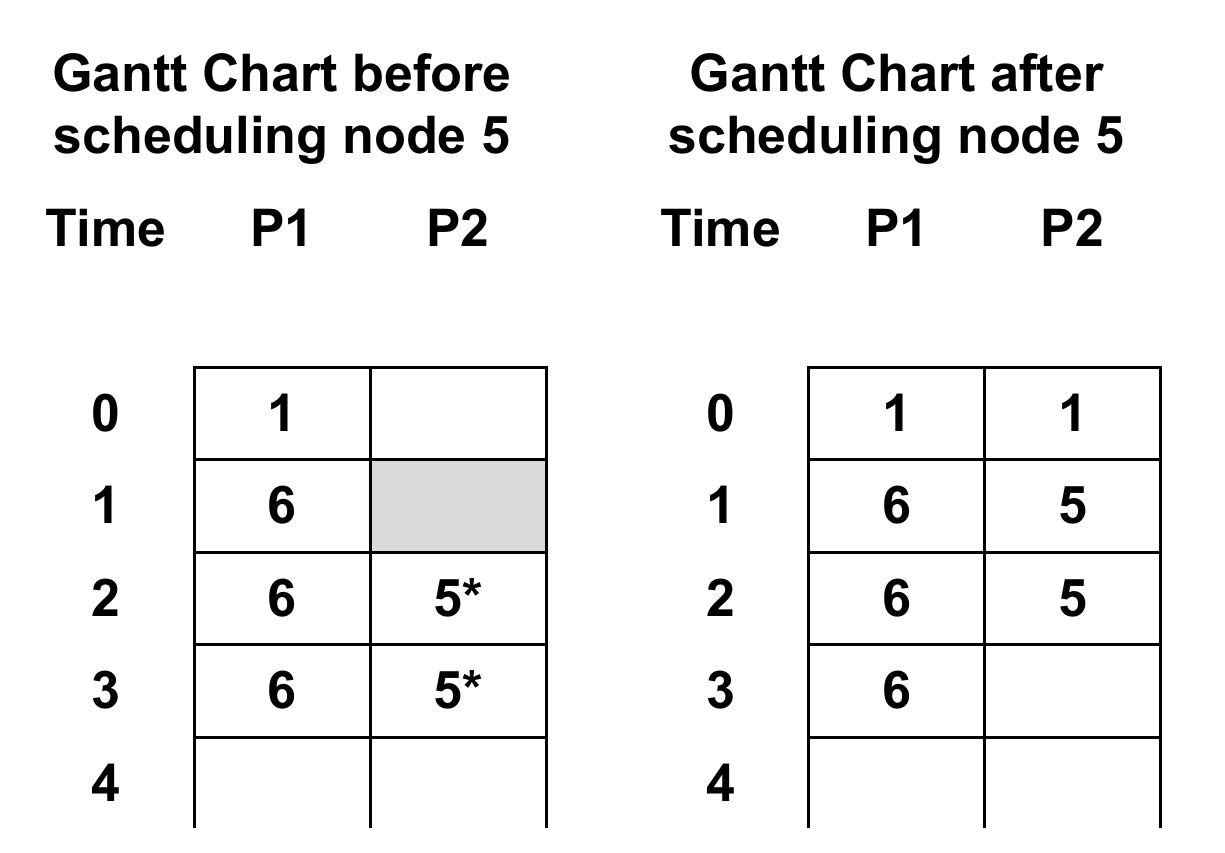}
    \caption{Scheduling of the node 5 using DSH. A "*" means that the node is not yet scheduled and a gray background is a delay induced by the communication between cores.}
    \label{fig:dsh}
\end{figure}

An example of this process is shown in Figure \ref{fig:dsh}. During the scheduling of node 5 on P2, idle time occurs due to communication from node 1 on P1, as shown in the rightmost chart of the figure. To reduce it, a copy of node 5's parent is placed on P2. This duplication, as illustrated in the leftmost chart, reduces both the delay and the task's start time.

\subsection{Exploring the solution space}

Another approach, proposed by Chou and Chung \cite{CHOU1994463}, explores the solution space directly using an algorithm tailored for scheduling. This method utilizes two relations between nodes in a graph $(V,E,t,w)$ to prune the search space:
\begin{itemize}
    \item \textbf{Dominance ($u\mathcal{D}v$):} if $\mathcal{P}(v) \supseteq \mathcal{P}(u)$ and $\mathcal{S}(u) \supset \mathcal{S}(v)$.
    \item \textbf{Equivalence ($u\mathcal{E}v$):} if $\mathcal{P}(v) = \mathcal{P}(u)$ and $\mathcal{S}(u) = \mathcal{S}(v)$.
\end{itemize}

\begin{figure}[hbt]
    \centering
    \includegraphics[width=0.8\linewidth]{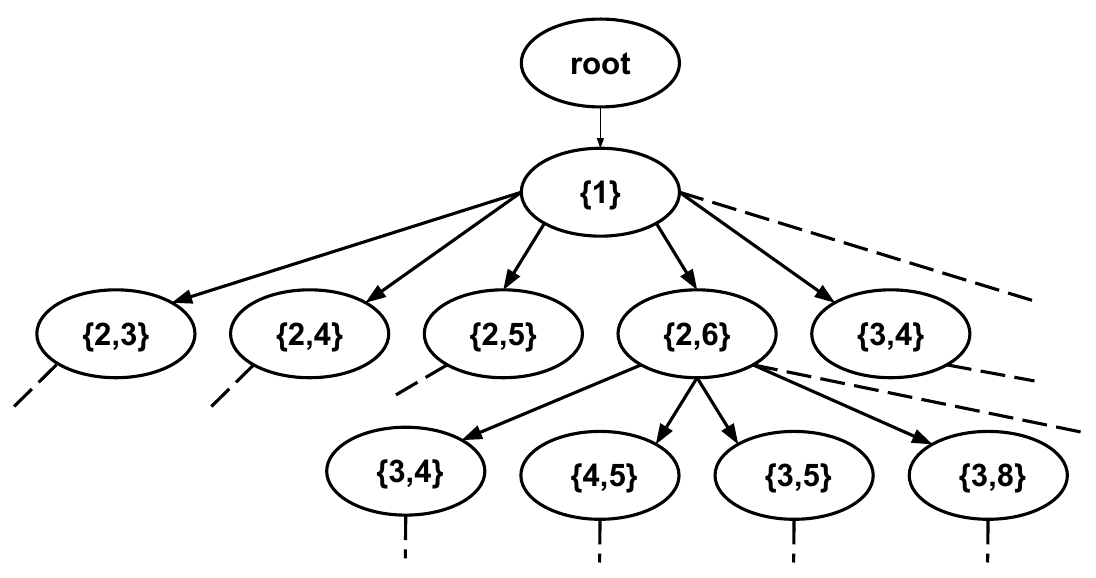}
    \caption{Partial solution tree for scheduling the graph in Figure \ref{fig:dag} on two cores}
    \label{fig:solution_tree}
\end{figure}

Similar to the pruning mechanisms in ILP solvers, these relations allow the algorithm to discard sub-optimal branches in a solution tree. Each \textit{S-node} in this tree represents a schedule state containing at most $m$ tasks (see Figure~\ref{fig:solution_tree}). Chou and Chung proved that if an \textit{S-node} dominates or is equivalent to another, the latter can be pruned without losing optimality. The optimal schedule is then identified by finding the shortest path from the root to the leaves of the pruned tree.

\section{Evaluation of the 
offline schedules}
\label{sec-evaluation}
We focus on the ILP representation and the scheduling heuristics. We use several sets of randomly generated DAGs and evaluate the methods on two main metrics: the speedup and the average computation time.

\subsection{Setup}
\textbf{Applications.} 
The evaluation uses three sets of randomly generated DAGs containing 20, 50, and 100 nodes, respectively. Each node is assigned a Worst-Case Execution Time and each edge a communication weight, both uniformly sampled from $[1, 10]$. 

The generation follows a three-step process: (1) node instantiation with unique indices; (2) edge creation by connecting lower-indexed nodes to higher-indexed ones to ensure acyclicity; and (3) a verification step to ensure a single sink node, as detailed in Section~\ref{application_model}.

The connectivity of these graphs is defined by their density, representing the ratio of actual edges $|E|$ to the maximum possible edges in a DAG:
\begin{equation}
    \text{density} = \frac{|E|}{|V|(|V|-1)/2}
\end{equation}
All test sets were generated with a density of 10\%, representing a moderately connected network.

\noindent\textbf{Metrics.}
The performance is evaluated using speedup and average computation time. Speedup is defined as the ratio between the single-core and multi-core makespans:
\begin{equation}
    \text{speedup} = \frac{\text{makespan single-core}}{\text{makespan multi-core}}
\end{equation}
where $C$ represents the makespan. The average computation time is calculated as the arithmetic mean of the durations required by a method to reach a solution across all test cases.

\begin{figure*}[hbt]
    \centering
    \subfloat[Speedup ISH results \label{fig:ish_speedup}]{\includegraphics[width=0.45\linewidth]{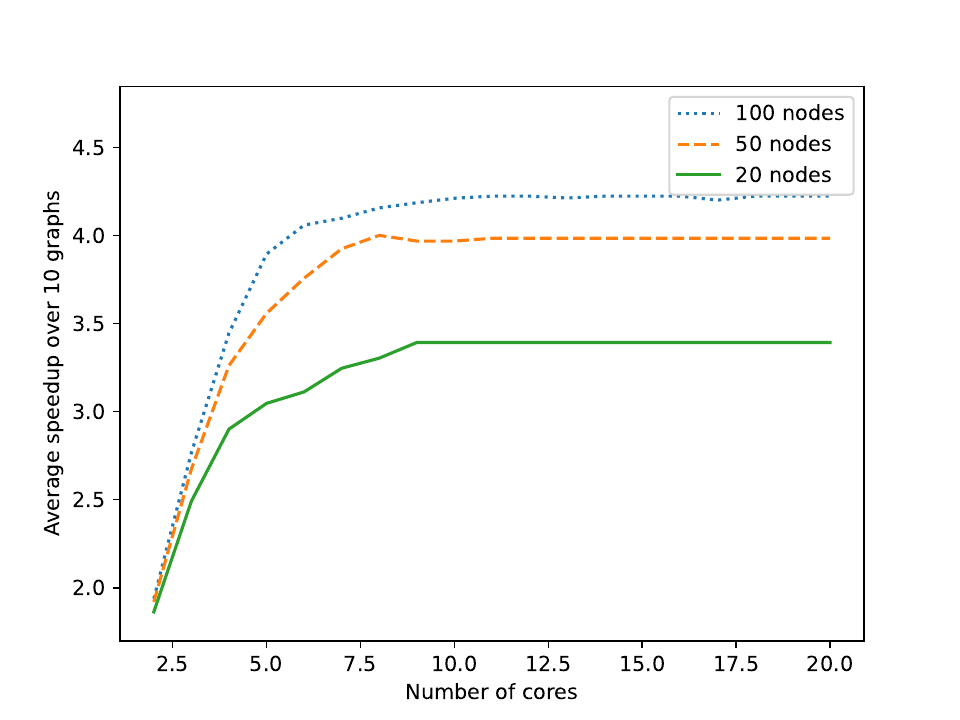}}
    \subfloat[Speedup DSH results \label{fig:dsh_speedup}]{\includegraphics[width=0.45\linewidth]{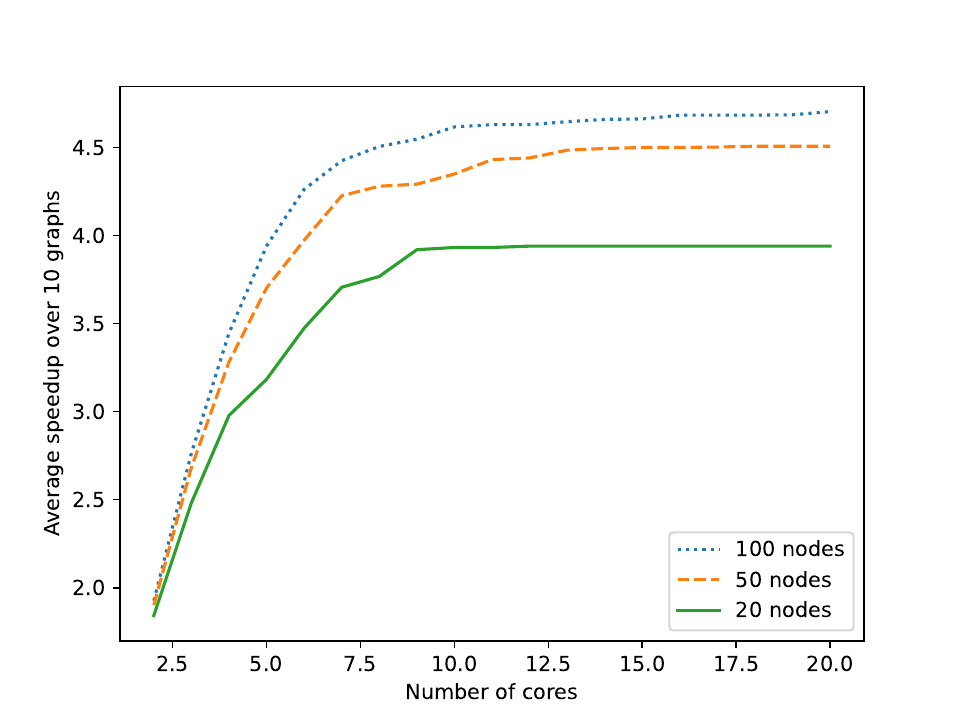}}\\
    \subfloat[Computation time ISH results \label{fig:ish_time}]{\includegraphics[width=0.45\linewidth]{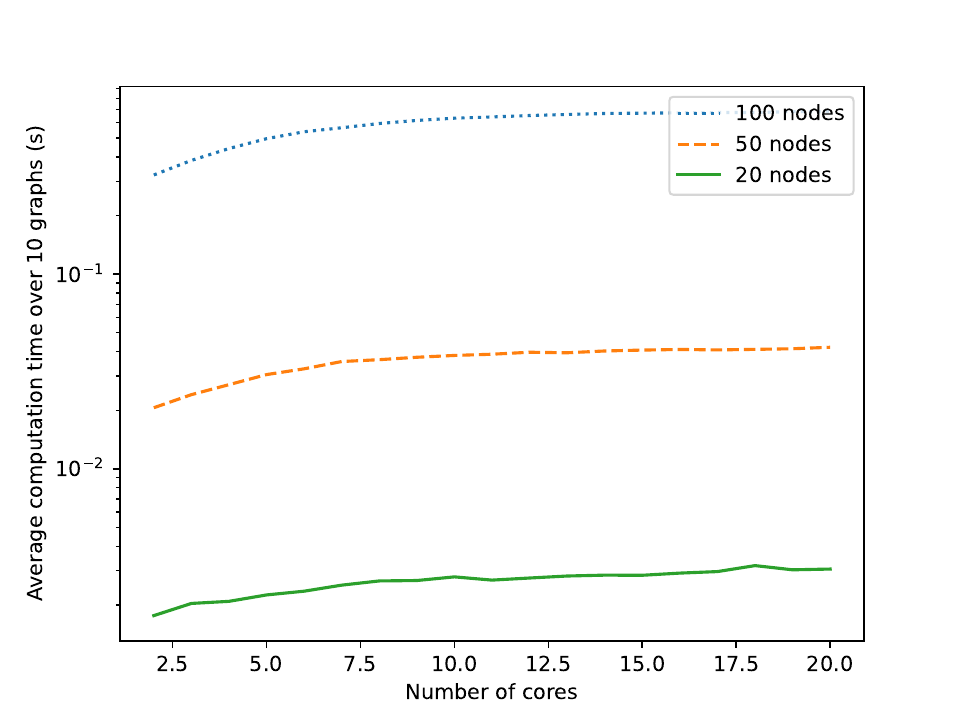}}
    \subfloat[Computation time DSH results \label{fig:dsh_time}]{\includegraphics[width=0.45\linewidth]{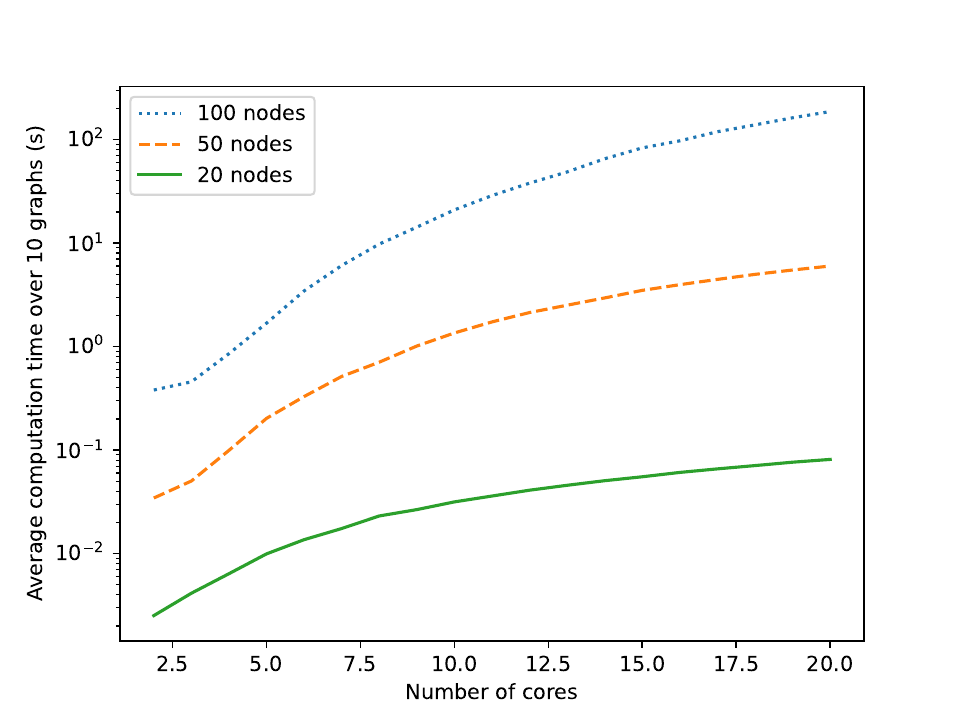}}
    \caption{Metrics  as a function of the number of cores}
\end{figure*}

\subsection{ISH and DSH evaluation}
This part evaluates the influence of the number of cores on both scheduling heuristics presented in subsection \ref{sh}. 
Figure \ref{fig:ish_speedup} (respectively \ref{fig:dsh_speedup}) represents the speedup provided by the ISH (respectively DSH) algorithm for a number of cores ranging from 2 to 20.

\noindent\textbf{Observation 1: maximal parallelism}
In both figures, the speedup grows until it reaches a plateau. 
The plateau is induced by the maximal parallelization value equal to the maximum number of parallel branches. For example, in Figure \ref{fig:dag}, this value would be 5. Having more cores than this number will not provide any additional benefits for the speedup. 

\noindent\textbf{Observation 2: highest speedup}
Under similar conditions, the DSH algorithm provides a higher or equal speedup than the ISH. 
This behavior has been observed in (\cite{TANG2020115} and \cite{Kruatrachue}): DSH is supposed to be slower but closer to the optimal solution, while ISH is faster but can only reach a near-optimal solution. 

Finally, we can see that the higher the number of nodes, the greater the speedup; thus, the use of parallelization becomes more advantageous for the corresponding application.


\noindent\textbf{Observation 3: faster computation}
The computation time results vary significantly between the two heuristics. As shown in Figures~\ref{fig:ish_time} and \ref{fig:dsh_time}, ISH is consistently faster and more stable, maintaining a steady order of magnitude for a given node count. Conversely, DSH execution time increases by one to two orders of magnitude as the number of cores grows, reaching nearly two minutes per graph. While still practical, DSH is significantly outperformed by ISH in terms of computational efficiency.

\noindent\textbf{Observation 4: memory footprint} 
By definition, DSH can duplicate nodes, creating an overhead in the memory footprint. 


\subsection{ILP evaluation}
To evaluate the ILP representation, we implemented the constraints using IBM's Optimization Programming Language (OPL) \cite{VanHentengry1999}. The problem is then solved using IBM's ILOG CP Optimizer \cite{Laborie2018}, run on a 24-core Intel(R) Xeon(R) CPU E5-2420 0 @ 1.90 GHz with 12 GB of RAM. To accelerate the tests, a timeout of 1 hour is set for each run. When the solver reaches this timeout, it returns the best solution found during the allocated time but can not assure that it is the optimal one.

\begin{figure*}[hbt]
    \centering
    \subfloat[Speedup  \label{fig:opl_speedup_3dvar}]{\includegraphics[width=0.4\linewidth]{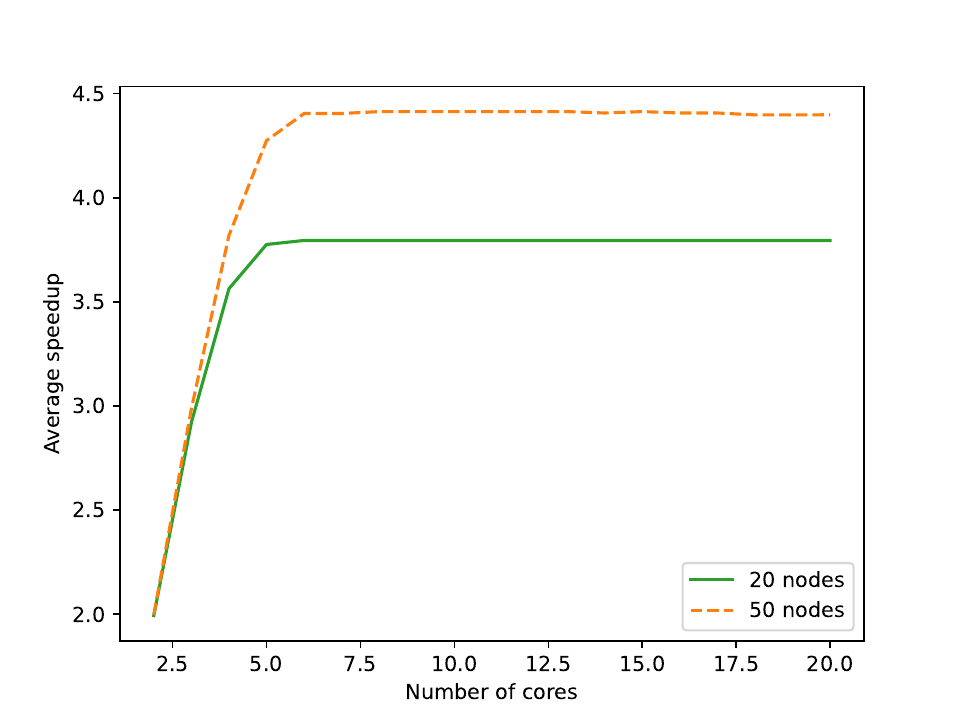}}
    \subfloat[Computation time \label{fig:opl_time_3dvar}]{\includegraphics[width=0.4\linewidth]{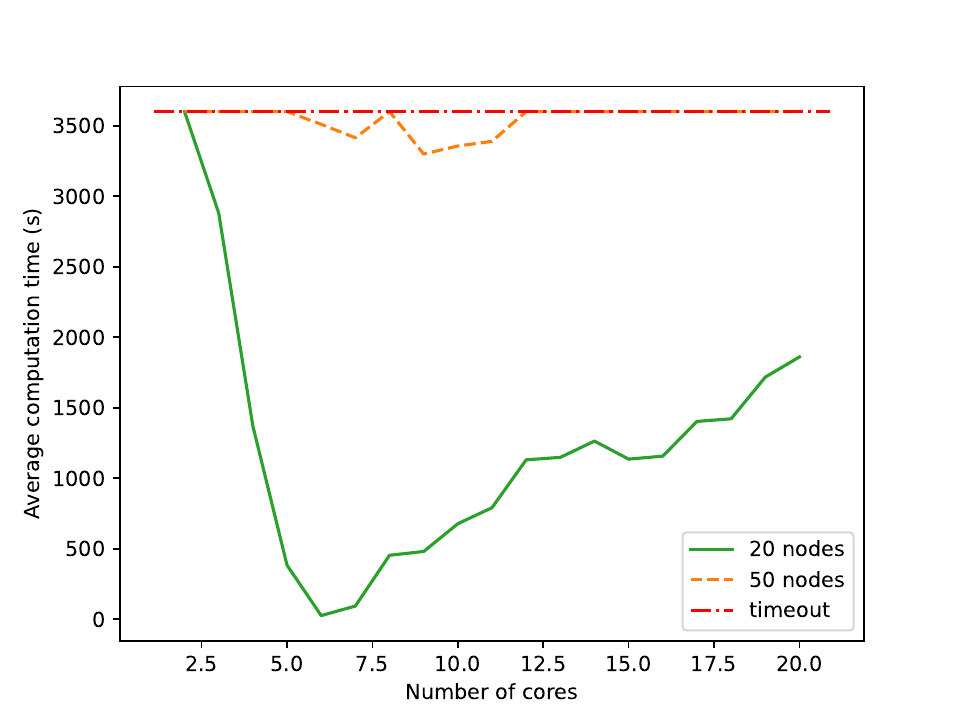}}
    \caption{Metrics evaluations of the modified ILP representation using OPL solver as a function of the number of core}
    \label{fig:3dvar}
\end{figure*}

The results are presented in Figure \ref{fig:3dvar}, with the speedup evaluation on the right and the computation time on the left for a number of cores ranging from 2 to 20. 
Solely the datasets of size 20 and 50 are represented here, as the timeout was too tight for the solver to find coherent solutions with larger graphs.

\noindent\textbf{Observation 1: comparison between Tang \textit{et al.} \cite{TANG2020115} and our optimized encoding}
When testing Tang's representation under similar conditions, the solver was not able to find solutions before the timeout, while ours returns at least one solution in all tested configurations. When returned, the schedule produced by our method is at least as good as the one returned by Tang's. Both representations yield similar performance in terms of speedup; but when the timeout is reached, ours allows the solver to explore the solution space further, and thus to produce better results in the same amount of time.

\noindent\textbf{Observation 2: optimal solution}
As for the previous heuristics, the speedup reaches a plateau after a few cores. For both datasets, the plateau is at a similar value than the one of DSH. 
The plateau is reached around 5 cores, while DSH needed 7 cores. 
This confirms that DSH only returns a near-optimal solution.

\noindent\textbf{Observation 3: slower computation}
The computation time quickly increases with the size of the evaluated graph. For 50-node graphs, the average computation time never goes below 54 minutes and often reaches the timeout. 

For a higher number of cores, both DSH and the ILP solver have similar speedup performance, with the heuristic being significantly faster (up to 10 times in our evaluations). However, for a lower number of cores, the solver outperforms the heuristic by finding optimal solutions for our problem, albeit still with a longer computation time. It is important to note that the graphs evaluated here are quite small compared to some models found in the industry, where the number of nodes easily exceeds 50 for modern networks. To limit the explosion of computation time, a hybrid solution should be considered, where a call to DSH gives a first schedule, which is then used as a starting point by the solver to find an optimal solution.

\section{Implementation in \acetone}
\label{sec-acetone-extension}
\subsection{Current state of the software}

\begin{figure}[hbt]
    \hspace{-.6cm}
    \includegraphics[width=1.15\linewidth]{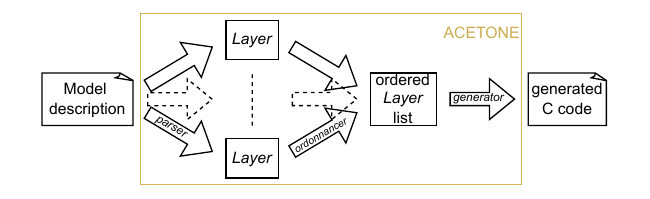}
    \caption{\acetone's workflow}
    \label{fig:workflow}
\end{figure}

The \acetone workflow is summarized in Figure \ref{fig:workflow}. Starting from a model description (NNet, ONNX, H5, or JSON), a parser extracts the network layers and creates the corresponding objects in the \acetone internal representation,
capturing for each layer the underlying operation and its relation with other layers (producers of its operands, or consumers of its output).
The scheduler arranges them topologically; the scheduler ensures each layer is visited after its parents have produced their output. This ordered list is passed to the code generator, which traverses it and produces the corresponding C implementation.

Each layer in \acetone is represented by a \textit{Layer} Python object associated with an implementation in C (a default template is provided to generate the implementation, with the possibility for the user to implement their own).  
These implementations are then printed sequentially in an \textit{inference function}, following the ordered list, with glue code passing the output of a layer as inputs of its children. 
As an example, Algorithm~\ref{algo:inference} provides an extract of the code generated for the LeNet-5 described in Figure~\ref{fig:lenet_sep}.
The external input is loaded in a local variable using the \textit{Input} layer at lines 3-4. From line 6 to 30, each layer is executed sequentially, the result being stored in statically allocated output variables, which are then referenced as inputs in the subsequent layer. Finally, an \textit{Output} layer, lines 32-33, retrieves the value computed from the last layer and puts it in the global output variable.

\begin{algorithm}[hbt]
    \caption{Inference function of LeNet5 Figure \ref{fig:lenet_sep}}
    \label{algo:inference}
    \begin{algorithmic}[1]
        \Function{inference}{float** inputs, float** outputs}
            \LComment{\color{blue} Input layer}
            \For{int i=0; i$<$nb\_input\_elements; ++i}
                \State $next\_layer\_in[i] = inputs[i];$
            \EndFor
            \State
            \LComment{\color{blue} Split layer}
            \State \textit{layer\_out\_1, layer\_out\_2 = split(next\_layer\_in);}
            \For{int i=0; i$<$nb\_layer\_elements; ++i}
                \State $next\_layer\_in\_top\_branch[i] = layer\_out\_1[i];$
                \State $next\_layer\_in\_bot\_branch[i] = layer\_out\_2[i];$
            \EndFor
            \State
            \LComment{\color{blue} Top branch Conv1 layer}
            \State \textit{layer\_out\_top = conv(next\_layer\_in\_top\_branch);}
            \For{int i=0; i$<$nb\_layer\_elements; ++i}
                \State $next\_layer\_in\_top\_branch[i] = layer\_out\_top[i];$
            \EndFor
            \State
            \LComment{\color{blue} Bottom branch Conv1 layer}
            \State \textit{layer\_out\_bot = conv(next\_layer\_in\_bot\_branch);}
            \For{int i=0; i$<$nb\_layer\_elements; ++i}
                \State $next\_layer\_in\_bot\_branch[i] = layer\_out\_bot[i];$
            \EndFor
            \State \textbf{\ldots}
            \State
            \LComment{\color{blue} Concat layer}
            \State \textit{layer\_out = concat(\\ next\_layer\_in\_top\_branch,\\ next\_layer\_in\_bot\_branch,\\)}
            \For{int i=0; i$<$nb\_input\_elements; ++i}
                \State $next\_layer\_in[i] = layer\_out[i];$
            \EndFor
            \State \textbf{\ldots}
            \State
            \LComment{\color{blue} Dense3 layer}
            \State \textit{layer\_out = dense(next\_layer\_in);}
            \LComment{Output layer}
            \For{int i=0; i$<$nb\_layer\_elements; ++i}
                \State $outputs[i] = layer\_out[i];$
            \EndFor
        \EndFunction
    \end{algorithmic}
\end{algorithm}

\subsection{Synchronization mechanisms}
\label{sec-synchronization-mechanisms}
The communication between nodes executed on different cores must ensure 
precedence between the production of an output and its consumption is respected.
The communication starts after the end of the producing layer and ends before the start of the consumer ones. Since we consider bare-metal setting, we implement the synchronization using the specificity of our platform: the UMA. 
For each pair of source and destination cores, 
a corresponding flag and array are reserved in the shared memory; the same flag and array will be used for different data the provided source and destination are the same. 
The sender only writes data into the array, while the receiver only reads and copies it into local buffers.
The flag ensures that only the intended receiver accesses the right data at the right time. Each data written to the array is uniquely identified by its sequence number. 
Before each communication, the flag's value is checked to ensure the sender does not overwrite data that has yet to be handled, and conversely the receiver waits for the expected data. If not, the core waits for the authorization. The data handling is then executed and the flag's value is incremented by one.


Thus, on an $m$-core target, we introduce at most $2m(m-1)$ variables to handle communication ($m(m-1)$ flags and $m(m-1)$ arrays). For a small number of cores, this remains limited, but it quickly exceeds acceptable values. For 20 cores, which is the maximum number of cores on which we evaluated our heuristics in Section \ref{sec-evaluation}, we need to introduce \textbf{760} new variables to manage the communication. In the industry, due to certification costs, safety-critical embedded systems rarely use more than four cores simultaneously. In such a configuration, only \textbf{24} new variables are added, which is easily manageable.

Note that the proposed solution is a trade-off between the memory and time requirements for the synchronization primitives. The use of a single buffer per core-to-core communication channel does limit the memory footprint of buffers, over the allocation of a buffer per communication. However it might introduce additional synchronization time for a writer waiting on a reader to read older data in the buffer. We are currently investigating alternative schemes to support non-blocking writes considering guarantees on the liveliness of data in communication channels, e.g. considering the best and worst producer and consumer arrivals for each communication.

\subsection{\acetone extension}
Using the \acetone sequential scheduler, all the layers are executed sequentially and share input and output variables when on the same branch. 
To enable parallelism, 
we update the \acetone scheduler to allocate layers to the available cores on the target, and generate a sub-schedule per core. The scheduler thus generates a separate list of layers per core, with additional layers inserted to capture outgoing or incoming communications.
Applying the generator directly to each list produces a set of independent \textit{inference functions}, one for each core, that can be run in parallel.

To represent inter-core communication, we introduce \textit{Writing} and \textit{Reading} operators. Based on the protocol in Section \ref{sec-synchronization-mechanisms}, the \textit{Writing} operator waits for a specific flag value, copies data to a shared variable, and updates the flag. Conversely, the \textit{Reading} operator waits for the flag update before copying data locally and increments the flag. Synchronization depends on flag values derived from the source and destination core schedules.

        


Algorithms \ref{algo:inference_core_0} and \ref{algo:inference_core_1} illustrate the generated code for the LeNet-5 networked showed in Figure \ref{fig:lenet_sep}, with a schedule using two cores without duplications. In this configuration, all the layers are executed on the first core (Algorithm \ref{algo:inference_core_0}), with the exception of the bottom convolution and pooling which are executed on the second core (Algorithm \ref{algo:inference_core_1}). 

\begin{algorithm}[hbt]
    \caption{Generated inference for the first core }
    \label{algo:inference_core_0}
    \begin{algorithmic}[1]
        \Function{inference\_1}{float** inputs, float** outputs}
            \LComment{\color{blue} Input layer}
            \For{int i=0; i$<$nb\_input\_elements; ++i}
                \State $next\_layer\_in[i] = inputs[i];$
            \EndFor
            \State
            \LComment{\color{blue} Split layer}
            \State \textit{layer\_out\_1, layer\_out\_2 = split(next\_layer\_in);}
            \For{int i=0; i$<$nb\_layer\_elements; ++i}
                \State $next\_layer\_in\_top\_branch[i] = layer\_out\_1[i];$
                \State $next\_layer\_in\_bot\_branch[i] = layer\_out\_2[i];$
            \EndFor
            \State
            \LComment{\color{red} Writing layer}
            \LComment{Waiting for the flag to be in writing state}
            \While{$flag\_0\_1 \neq 0$}
            \EndWhile
            \LComment{Loading the data in the communication variable $comm$}
            \For{int i=0; i$<$comm\_size; ++i}
                \State $comm\_0\_1[i] = next\_layer\_in\_bot\_branch[i];$
            \EndFor
            \LComment{Setting the flag to reading state}
            \State $flag\_0\_1 += 1$
            \State
            \LComment{\color{blue} Top branch Conv1 layer}
            \State \textit{layer\_out\_top = conv(next\_layer\_in\_top\_branch);}
            \For{int i=0; i$<$nb\_layer\_elements; ++i}
                \State $next\_layer\_in\_top\_branch[i] = layer\_out\_top[i];$
            \EndFor
            \State \textbf{\ldots}
            \State
            \LComment{\color{red}Reading layer}
            \While{$flag\_1\_0 \neq 1$}
            \EndWhile
            \LComment{The data can be read in $comm$}
            \For{int i=0; i$<$comm\_size; ++i}
                \State $next\_layer\_in\_bot\_branch[i] = comm\_1\_0[i];$
            \EndFor
            \State $flag\_1\_0 += 1$
            \State
            \LComment{Concat layer}
            \State \textbf{\ldots}
        \EndFunction
    \end{algorithmic}
\end{algorithm}

\begin{algorithm}[hbt]
    \caption{Generated inference for the second core }
    \label{algo:inference_core_1}
    \begin{algorithmic}[1]
        \Function{inference\_2}{float** inputs, float** outputs}
            \LComment{\color{red} Reading layer}
            \State
            \While{$flag\_0\_1 \neq 1$}
            \EndWhile
            \LComment{The data can be read in $comm$}
            \For{int i=0; i$<$comm\_size; ++i}
                \State $next\_layer\_in\_bot\_branch[i] = comm\_0\_1[i];$
            \EndFor
            \State $flag\_0\_1 += 1$
            \State
            \LComment{\color{blue} Top branch Conv1 layer}
            \State \textit{layer\_out\_top = conv(next\_layer\_in\_top\_branch);}
            \For{int i=0; i$<$nb\_layer\_elements; ++i}
                \State $next\_layer\_in\_top\_branch[i] = layer\_out\_top[i];$
            \EndFor
            \State \textbf{\ldots}
            \LComment{\color{red} Writing layer}
            \LComment{Waiting for the flag to be in writing state}
            \While{$flag\_1\_0 \neq 0$}
            \EndWhile
            \LComment{Loading the data in the communication variable $comm$}
            \For{int i=0; i$<$comm\_size; ++i}
                \State $comm\_1\_0[i] = next\_layer\_in\_bot\_branch[i];$
            \EndFor
            \LComment{Setting the flag to reading state}
            \State $flag\_1\_0 += 1$
        \EndFunction
    \end{algorithmic}
\end{algorithm}

Two synchronization points manage data transfer between cores: from \textit{Split} (Core 0) to convolution (Core 1), and from \textit{Pool} (Core 1) to \textit{Concat} (Core 0). Each utilizes paired \textit{Writing} and \textit{Reading} operators with shared flags. In the first instance, Core 1 waits at a synchronization barrier (Alg. \ref{algo:inference_core_1}, l.4) until Core 0 finishes its layer and updates the communication flag (Alg. \ref{algo:inference_core_0}, l.19). This update triggers Core 1 to copy the data and release the flag. The cores then proceed in parallel until the second synchronization, which executes this protocol in reverse.

\subsection{\textit{Layers}'s WCET evaluation}
\label{sec-wcet-eval}
We use the tool OTAWA \cite{otawa} to evaluate the WCET of the 
implemented \textit{Layers} with \acetone. As described in \cite{dealbuquerquesilva_et_al:LIPIcs.ECRTS.2022.3}, OTAWA does not support the micro-architecture of our hardware target. Thus, we compile the code for \textit{lpc2138} ARM-based architecture. The difference in targets prevents us from comparing the predicted WCET and the experimental results. Our objective in this section is to establish an evaluation of the gain in total WCET given by the parallelization process.

\begin{figure}[hbt]
    \centering
    \includegraphics[width=\linewidth]{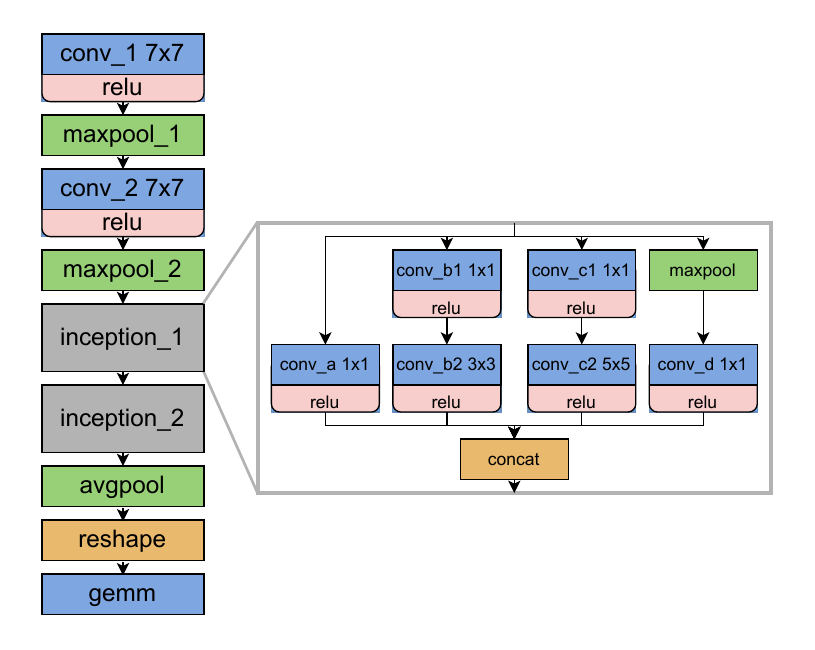}
    \caption{Example of a Googlenet architecture}
    \label{fig:googlenet}
\end{figure}

To have an evaluation, we assessed the network presented in Figure \ref{fig:googlenet}, which is based on the GoogleNet architecture \cite{szegedy2014goingdeeperconvolutions}. We chose this architecture for the Inception modules, illustrated in the right box of the figure. The module features four independent branches.

We estimated the WCET of each layer and compiled the results in Table~\ref{tab:wcet_layers}. The reshape operation impacts only the dimensions of the input without changing the order of its elements. In the generated code, each tensor is encoded with a 1D array and  reshaping a 1D tensor does not modify anything, leading to a zero WCET.

\begin{table}[hbt]
\centering
\begin{tabular}{l|c}
\textbf{Layer Name}   & \textbf{WCET [cycles]} \\ \hline
Input Layer           & $5.27 \times 10^6$ \\
conv\_1               & $8.16 \times 10^9$ \\
maxpool\_1            & $1.22 \times 10^8$ \\
conv\_2               & $1.59 \times 10^{10}$ \\
maxpool\_2            & $2.71 \times 10^7$ \\
inception\_1/conv\_a  & $4.57 \times 10^8$ \\
inception\_1/conv\_b1 & $2.86 \times 10^8$ \\
inception\_1/conv\_b2 & $7.92 \times 10^8$ \\
inception\_1/conv\_c1 & $5.72 \times 10^7$ \\
inception\_1/conv\_c2 & $1.63 \times 10^8$ \\
inception\_1/maxpool  & $2.49 \times 10^7$ \\
inception\_1/conv\_d  & $2.29 \times 10^8$ \\
inception\_1/concat   & $6.06 \times 10^6$ \\
inception\_2/conv\_a  & $6.86 \times 10^8$ \\
inception\_2/conv\_b1 & $3.43 \times 10^8$ \\
inception\_2/conv\_b2 & $1.14 \times 10^9$ \\
inception\_2/conv\_c1 & $8.58 \times 10^7$ \\
inception\_2/conv\_c2 & $2.53 \times 10^8$ \\
inception\_2/maxpool  & $2.49 \times 10^7$ \\
inception\_2/conv\_d  & $2.29 \times 10^8$ \\
inception\_2/concat   & $7.49 \times 10^6$ \\
avgpool               & $2.51 \times 10^6$ \\
reshape               & 0 \\
gemm                  & $2.67 \times 10^7$ \\
Output Layer          & $3.51 \times 10^4$ \\ \hline
\textbf{Total Sum}    & $\mathbf{2.90 \times 10^{10}}$ \\
\end{tabular}
\caption{WCET bounds produced by OTAWA for each individual layer.}
\label{tab:wcet_layers}
\end{table}

\begin{figure}[hbt]
    \centering
    \includegraphics[width=\linewidth]{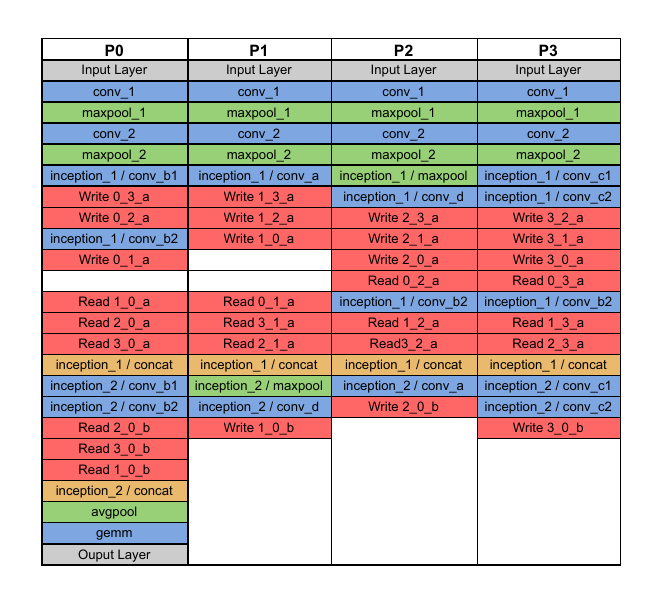}
    \caption{Googlenet scheduling on four cores}
    \label{fig:googlenet-scheduling}
\end{figure}

Using Table \ref{tab:wcet_layers}, the network is scheduled on four cores by the DSH heuristic as shown in Figure \ref{fig:googlenet-scheduling}. Each communication, highlighted in red in the figure, is named using the following norm: \textit{source\_destination\_identifier}. For example, the communication $2\_0\_b$ is the data transfer named $b$ going from core $2$ to core $0$.

To handle those communications, operators have been added to the program and need to be taken into account. We estimated the WCET of their data handling part and compiled it in Table \ref{tab:wcet_sync}. \textit{Writing} and \textit{Reading} operators having similar code, their WCET is the same at each end of a communication.

\begin{table}[hbt]
\centering
\begin{tabular}{l|c}
\textbf{Communication Name}    & \textbf{WCET [cycles]} \\ \hline
2\_0\_b                        & $3.58 \times 10^5$ \\
0\_1\_a                        & $2.98 \times 10^5$ \\
1\_Y\_a                        & $2.38 \times 10^5$ \\
0\_2\_a, 0\_3\_a               & $1.49 \times 10^5$ \\
1\_0\_b, 2\_Y\_a, 3\_Y\_Z      & $1.19 \times 10^5$ \\ \hline
\end{tabular}
\caption{WCET bounds produced by OTAWA for the synchronization layers ($Y$ denotes any core and $Z$ denotes any name).}
\label{tab:wcet_sync}
\end{table}

The global WCET is constructed layer-by-layer using Table \ref{tab:wcet_layers}, synchronizing cores at each barrier by adopting the maximum accumulated WCET. The parallelized model achieves $2.68 \times 10^{10}$ cycles, an 8\% gain over the sequential $2.90 \times 10^{10}$ cycles. This modest overall speedup is due to the large contributions of sequential layers \texttt{conv\_1} and \texttt{conv\_2} to the WCET. However, the parallelizable segment (\texttt{maxpool\_2} to \texttt{inception\_2/concat}) shows a 46\% gain, reducing execution from $4.81 \times 10^9$ to $2.60 \times 10^9$ cycles.

\subsection{Evaluation on the target}

The generated code is then run on a Texas Instruments Keystone II SoC \cite{keystone}, which is composed of a cluster of four ARM Cortex-A15 cores implementing the ARMv7 architecture. We evaluated the number of cycles taken by each layer's implementation to execute in both sequential and parallelized configurations, and compiled the results in Table \ref{tab:measured_wcet}. When several instances of the same layer exist on different cores, only the highest computation time is represented.

\begin{table}[hbt]
\centering
\begin{tabular}{l|c|c}
\textbf{Layer name} & \textbf{Single-core} & \textbf{Multi-core} \\ \hline
Input Layer           & $9.75 \times 10^5$ & $3.34 \times 10^6$ \\
conv\_1               & $6.92 \times 10^8$ & $6.86 \times 10^8$ \\
maxpool\_1            & $1.26 \times 10^7$ & $1.32 \times 10^7$ \\
conv\_2               & $1.45 \times 10^9$ & $1.45 \times 10^9$ \\
maxpool\_2            & $2.61 \times 10^6$ & $2.62 \times 10^6$ \\
inception\_1/conv\_a  & $1.36 \times 10^7$ & $1.37 \times 10^7$ \\
inception\_1/conv\_b1 & $8.46 \times 10^6$ & $8.63 \times 10^6$ \\
inception\_1/conv\_b2 & $6.29 \times 10^7$ & $7.60 \times 10^7$ \\
inception\_1/conv\_c1 & $7.53 \times 10^6$ & $1.86 \times 10^6$ \\
inception\_1/conv\_c2 & $1.16 \times 10^7$ & $1.19 \times 10^7$ \\
inception\_1/maxpool  & $2.55 \times 10^6$ & $2.49 \times 10^6$ \\
inception\_1/conv\_d  & $6.96 \times 10^6$ & $6.94 \times 10^6$ \\
inception\_1/concat   & $4.37 \times 10^5$ & $4.56 \times 10^5$ \\
inception\_2/conv\_a  & $2.03 \times 10^7$ & $2.04 \times 10^7$ \\
inception\_2/conv\_b1 & $1.01 \times 10^7$ & $1.02 \times 10^7$ \\
inception\_2/conv\_b2 & $9.48 \times 10^7$ & $9.53 \times 10^7$ \\
inception\_2/conv\_c1 & $2.54 \times 10^6$ & $2.62 \times 10^6$ \\
inception\_2/conv\_c2 & $1.76 \times 10^7$ & $1.92 \times 10^7$ \\
inception\_2/maxpool  & $2.55 \times 10^6$ & $2.62 \times 10^6$ \\
inception\_2/conv\_d  & $6.90 \times 10^6$ & $6.94 \times 10^6$ \\
inception\_2/concat   & $1.02 \times 10^6$ & $5.29 \times 10^5$ \\
avgpool               & $1.69 \times 10^5$ & $1.42 \times 10^5$ \\
reshape               & 0                  & 0                  \\
gemm                  & $2.67 \times 10^6$ & $2.69 \times 10^6$ \\
Output Layer          & $3.22 \times 10^3$ & $3.77 \times 10^3$ \\ \hline
\textbf{Total}        & $\mathbf{2.42 \times 10^9}$ & $\mathbf{2.22 \times 10^9}$
\end{tabular}
\caption{Measured WCET [cycles] for the different layers in both single-core and multi-core contexts.}
\label{tab:measured_wcet}
\end{table}

\noindent\textbf{Observation 1: Multi-core interference} 
For almost all layers, the same order of magnitude for the number of cycles is found in both cases. The only exceptions are the Input layer, where the multiple cores heavily interfere with each other, and inception\_2/concat. The measured WCET for the latter one is high, but this only occurred in one measure among the ten, with all the other ones being around 5.00e+5 cycles.

\noindent\textbf{Observation 2: Computation time distribution}
We observe a similar distribution of computation time as with the estimated WCET, with conv\_2 being the most demanding operation, closely followed by conv\_1.

\noindent\textbf{Observation 3: Parallelism gain}
We achieved a 8\% gain in acceleration, which corresponds to the theoretical gain found in Section \ref{sec-wcet-eval}. When focusing only on the highly parallelizable part of the model---from the beginning of the maxpool\_2 layer to the end of inception\_2/concat---we observe a gain of 31\% ($2.72 \times 10^8$ cycles in sequential execution versus $1.86 \times 10^8$ cycles in the parallel code). This difference can be explained by the check before the \textit{Writing} operation, which ensures that the data has been read before it is overwritten; this postpones the operation and increases the delay.

\section{Conclusion}
\label{sec-conclusion}
In this work, we have defined the scope of our problem and explored solutions from the literature to extend \acetone in order to generate parallel code and fully exploit the multi-core architecture of our systems. After reviewing the state of the art on DAG scheduling, we identified several resolution methods, conducted a study on the scalability of the selected heuristics, and compared their output schedules using normalized metrics.

We also focused on extending \acetone by creating new templates that allow the use of synchronization mechanisms and the repartition of tasks across several cores. These mechanisms are target-dependent, as they rely on specific interrupts that, when addressed to a given core, do not impact any other core, and on busy waiting by setting binary flags in shared memory and having the core stop until these flags are in the passing state. These mechanisms are necessary both for enforcing precedence constraints—by ensuring that a node is computed after all its parents on other cores—and for preserving the integrity of data in shared memory, by preventing simultaneous writing.

This extension is then verified using both a static WCET analysis with the dedicated tool OTAWA and experimental measurements on a Texas Instruments KeyStone platform.



For future work, we identified limitations in the paper to address, such as the strong constraints on the target (UMA, homogeneous cores, \ldots). We plan to extend the theory to take into account the use of non-homogeneous cores or accelerators included in the target, and to adapt both the scheduling heuristics and the ILP representation to this new problem. We also plan to explore other parallelization methods to improve the current implementation in \acetone.

\paragraph{Acknowledgement}
This work has benefited from the AI cluster ANITI2 and from the PATO project funded by the French
government through the France Relance program, based on the funding from and by the European Union
through the NextGenerationEU program.

\bibliographystyle{abbrv}
\PHMbibliography{erts}

\end{document}